\documentclass[twocolumn,letterpaper,aps,prd,longbibliography,superscriptaddress,nofootinbib,floatfix,amsmath,amssymb]{revtex4-1}

\usepackage{graphicx}	
\usepackage{xspace}     
\usepackage{subfigure}
\RequirePackage{lineno}
\usepackage[dvipsnames]{xcolor}
\usepackage{siunitx}  
\usepackage[normalem]{ulem}

\newcommand{\vc}[1]{\boldsymbol{#1}}

\begin{document}

\ \\[-2.0cm] {\begin{flushright} JLAB-THY-23-3782 \end{flushright}}


\title{Light quark and antiquark constraints from new electroweak data}

\author{A.~Accardi}
\affiliation{Hampton University, Hampton, VA, USA}
\affiliation{Jefferson Lab, Newport News, VA, USA}
\author{X.~Jing}
\affiliation{Southern Methodist University, Dallas, TX, USA}
\author{J.F.~Owens}
\affiliation{Florida State University, Tallahassee, FL, USA}
\author{S.~Park}\affiliation{Mississippi State University, Mississippi State, MS, USA}\affiliation{Center for Frontiers in Nuclear Science, Stony Brook, NY, USA} 

\date{\today}

\begin{abstract}
We present a new parton distribution function analysis which includes new data for W boson production in proton-proton collisions and lepton pair production in proton-proton and proton-deuteron collisions. The new data provide strong constraints on the light antiquark parton distribution functions in the proton. We identify an interesting correlation between the $d/u$ ratio and the $\bar{d}/\bar{u}$ ratio which leads to a modification of our previous results for the $d/u$ ratio as the parton momentum fraction $x \rightarrow 1.$

\end{abstract}
	
\maketitle

\section{Introduction}

Predictions for high energy lepton-hadron and hadron-hadron hard collisions rely on perturbative QCD-based calculations for the parton-parton scattering cross sections. These are then convoluted with the appropriate parton distribution functions (PDFs) to obtain predictions for experimentally measured observables. A recent evaluation of PDF determinations can be found in Ref.~\cite{Accardi:2016ndt}. Global fits for PDFs by the CJ Collaboration have focused on simultaneously extending the reach in $x$ towards $x\approx 1$ and reducing the minimum value of the squared four-momentum transfer, $Q^2$, included in the fitting process \cite{Owens:2012bv, Accardi:2016qay}. The focus of this analysis is, instead, on the flavor dependence of the light-quark sea, specifically the behavior in $x$ of the ratio $\bar d/\bar u.$ One important source of information on the behavior of this ratio is lepton pair production using a proton beam on both proton and deuteron targets. The most precise such data available previously came from the E866 experiment \cite{NuSea:1998kqi}. These data suggested that $\bar d/\bar u$ initially rose from a value of unity to a maximum near $x \approx 0.15$ followed by a fall-off to a value below unity by $x \approx 0.30$. However, by this region in $x$ the data were statistically limited. The \texttt{CJ12} PDFs \cite{Owens:2012bv} were parametrized in such a way that the ratio could follow the data to values below one. This led to a rapid fall-off of the $\bar d$ PDF below 0 as $x$ increased beyond about 0.3. The \texttt{CJ15} PDFs \cite{Accardi:2016qay} employed an alternative parametrization which constrained $\bar d/\bar u$ to approach one from above at large values of $x$. 
   
New data from the SeaQuest experiment \cite{SeaQuest:2021zxb}, the successor to the E866 experiment, have a greater reach in $x$ as well as increased statistics. Additionally, new data \cite{STAR:2020vuq} from the STAR Collaboration on $W$ boson production in proton-proton collisions have become available. These data also offer additional constraints on $\bar d/\bar u$. It is the purpose of this analysis to assess the effects of these new data sets on the behavior of $\bar d/\bar u$ over the $x$ range out to $x \approx 0.4$.
At the same time, we expose a little noticed correlation between the light-antiquark and light-quark ratios inherent in the available lepton pair production data and in the mid-rapidity weak boson production data that affects the extrapolation of the $d/u$ ratio to values of $x$ approaching 1.
In particular, we examine different parametrizations to see their effects on the extracted ratios. Preliminary results have been presented at DIS 2021 \cite{Park:2021kgf}, and analyses of the new data have also been performed by the CT \cite{Guzzi:2021fre} and JAM \cite{Cocuzza:2021cbi} collaborations.  
   
The plan of the paper is as follows. In Section~\ref{sec:CJ_framework} the framework for the global fits is described, including the parametrizations used for the various PDFs and the higher-twist and nucleon off-shell corrections. Section~\ref{sec:dataset} contains a discussion of the data sets used with special attention paid to the new data, while Section~\ref{sec:results} presents the results of this analysis. The conclusions are summarized in Section~\ref{sec:conclusions}.

\section{Light quarks and antiquarks in the CJ global analysis}
\label{sec:CJ_framework}

The new \texttt{CJ22} global fit we report in this paper combines elements of the \texttt{CJ15} \cite{Accardi:2016qay} and \texttt{CJ15-a} \cite{Accardi:2019ofk} analyses in order to provide sufficient flexibility in the determination of the mid-$x$ $\bar d/\bar u$ ratio after the inclusion of the STAR and SeaQuest data. The new fit will also allow us to properly analyze the correlation of the mid-$x$ $\bar d/\bar u$ ratio and the large-$x$ $d/u$ ratio induced by weak boson production data and how these impact the extrapolation of $d/u$ to $x \to 1$. In this section we focus on the methodological and numerical aspects of our fits, and in the next one we will discuss the global data set we utilize.

\subsection{PDF parametrization and theoretical setup}
\label{sec:PDF_par}

The latest CTEQ-JLab global fit (\texttt{CJ15}) was performed using the world deep-inelastic scattering (DIS) data set, as well as a variety of jet and electroweak boson production measurements \cite{Accardi:2016qay}. Among these, lepton pair production measurements by the E866 experiment at Fermilab provided the strongest constraints on the light antiquark sea, covering the $0.015 \lesssim x \lesssim 0.3$ parton momentum fraction region, with additional sensitivity provided by fixed target DIS data, in particular from the NMC experiment.

At the input scale of $Q^2_0 = 1.69$ GeV$^2$, a standard five-parameter functional form was used for most of parton species, including the $\bar u + \bar d$ combination: 
\begin{equation}
    xf(x, Q^2_0) = a_0 x^{a_1} (1-x)^{a_2}(1+a_3 \sqrt{x} + a_4 x) \ 
\label{eq:CJ15-par}
\end{equation}
The valence $d$ quark was however allowed to mix with the valence $u$ quark parametrization at large $x$, as to allow a finite limit for the $d/u$ ratio: 
\begin{align}
    d_v(x,Q_0^2) \to a_0^{d_v} \Big( \frac{d_v(x,Q_0^2)}{a_0^{d_v}} + b \, x^c \,  u_v(x,Q_0^2) \Big) \ ,
\label{eq:du_ratio}
\end{align}
with $b$ and $c$ as two additional parameters. As a result, the ratio $d_v/u_v$ could tend to a non-zero value as $x \to 1$, provided that $a_2^{d_v} > a_2^{u_v}$, which is usually the case. As in the \texttt{CJ15} and \texttt{CJ15-a} fits, only the $b$ parameter was left free and $c=2$ kept fixed, since the
new data do not provide additional constraints on the valence quark ratio at large $x$.

Turning to the light antiquarks, the $\bar d / \bar u$ ratio in the original \texttt{CJ15} fit was parametrized as
\begin{equation}
    \bar{d}/\bar{u} = a_0\,x^{a_1}(1-x)^{a_2} + 1 + a_3 x(1-x)^{a_4} 
\end{equation}
due to the limited $x$ coverage of the E866 data and the sharp downturn these required of the $\bar{d}/\bar{u}$ ratio. With this parametrization we also enforced the theoretical expectation from most modeling efforts that $\bar d/\bar u$ remains greater than or equal to one all the way up to $x \to 1$, and tends to 1 in that limit. This assumption was however revisited in Ref.~\cite{Accardi:2019ofk}, where the $\bar d - \bar u$ difference was considered instead of $\bar d / \bar u$ and parametrized as in Eq.~\eqref{eq:CJ15-par},
\begin{equation}
    x\big(\bar d-\bar u \big) = \bar a_0 \, x^{\bar a_1}
    (1-x)^{\bar a_2}(1 + \bar a_4 x) \ ,
\label{eq:CJ12-par}
\end{equation}
with the resulting fit called \texttt{CJ15-a}. Even if the new parametrization allowed for it, no strong indication of a sign change in the $\bar d - \bar u$ asymmetry in the $x \lesssim 0.3$ region measured by E866 was found.

With the new data from STAR sensitive to the smaller-$x$ rise of the $\bar d - \bar u$ asymmetry, and the new SeaQuest data constraining the $\bar d / \bar u$ ratio at $0.15 \lesssim x \lesssim 0.4$, well across the region where E866 indicated this would drop below 1, we can now revisit this whole issue. To allow for sufficient versatility in the description of the light quark sea, in this paper we will utilize the more flexible parametrization \eqref{eq:CJ12-par}. Furthermore, we will leave the $\bar a_2$ parameter free instead of fixing it to 2.5 units larger than the corresponding parameter for the $\bar u + \bar d$ combination as done in the \texttt{CJ15-a} analysis, thus providing additional freedom to the $\bar d / \bar u$ ratio in the limiting $x \to 1$ region.

Apart from this change in parametrization, we will adopt the same theoretical setup as in the \texttt{CJ15} fits, as described in Ref.~\cite{Accardi:2016qay}. In particular: we perform fits at next-to-leading order accuracy in the ACOT-$\chi$ heavy quark scheme; include target mass corrections for DIS data according to the OPE prescription by Georgi and Politzer \cite{Georgi:1976ve,DeRujula:1976baf}; and adopt the ``weak-binding approximation'' to correct for nucleon binding and Fermi motion in DIS and DY cross sections on deuteron targets, with the AV18 deuteron wave function describing the nucleon dynamics inside the target. Higher-twist corrections for DIS structure functions and off-shell nucleon corrections in deuteron targets will be discussed in more detail in the next subsection.

From a numerical point of view, and at variance with the \texttt{CJ15} family of analyses, NLO QCD corrections to the calculation of $W$ and $Z$ production cross sections were implemented by means of the APPLgrid \cite{Carli:2010rw} fast NLO interface. The necessary coefficient grids were calculated by means of the MCFM 6.8 event generator \cite{Campbell:1999ah,Campbell:2012uf}, and tested against the Tevatron weak boson production data already included in the \texttt{CJ15} analysis. Details about the grids for weak boson production at the Relativistic Heavy Ion Collider (RHIC) will be discussed in Section~\ref{sec:dataset}.

\subsection{Higher-twist and off-shell corrections}

In DIS at low $Q^2$ values, power suppressed corrections exist beyond target mass corrections, for example, genuine multiparton correlations; missing higher order perturbative corrections can also resemble power corrections at small scale values.  Regardless of their origin, we account for these residual power suppressed contributions by using a phenomenological multiplicative factor to modify the proton and nucleon structure functions as in all earlier CJ fits, 
\begin{align}
    F_2(x,Q^2)
        = F_2^{\rm LT}(x,Q^2)
        \left( 1 + \frac{C(x)}{Q^2} \right),
\end{align}
where $F_2^{\rm LT}$ denotes the leading twist structure function
including target mass corrections, and $C$ is assumed to be isospin independent due to the relatively weak constraining power of the adopted data sets \cite{Virchaux:1991jc,Alekhin:2003qq,Blumlein:2008kz}. 
Following common usage, we generically refer to the fitted
$1/Q^2$ term as a ``higher-twist'' (HT) correction, and parametrize the
coefficient function $C$ by
\begin{align}
    C(x) = a_\textsc{ht}\, x^{b_\textsc{ht}} (1+c_\textsc{ht} x) \, .  
\end{align}
For ease of notation, we collect the higher-twist parameters in the vector
\begin{align}
    \vc a_\textsc{ht} = (a_\textsc{ht},b_\textsc{ht},c_\textsc{ht}) \, .
\end{align}

In deuteron targets, the nucleons are off their $m_N^2$ mass shell with a four-momentum squared $p_N^2 \neq m_N^2$. While the off-shell nucleon PDF, $\widetilde{f}$, is not an observable per se, its dependence on the nucleon virtuality $p_N^2$ can be studied within a given theoretical framework. For weakly bound nucleons such as in the deuteron, for example, one may expand $\widetilde{f}$ to lowest order about its mass shell \cite{Kulagin:1994fz,Kulagin:1994cj},
\begin{align}
    \widetilde{f} (x,p_N^2,Q^2)
    &= f(x,Q^2)
        \left( 1 + \frac{p_N^2-M^2}{M^2} \delta f(x,Q^2) \right),
\label{eq:qoff}
\end{align}     
and the off-shell correction function $\delta f$ can be parametrized and fitted to data. In this work, we adopt the flavor-independent \texttt{CJ15} parametrization
\begin{align}
  \delta f(x) = {\cal N} (x-x_0) (x-x_1) (1+x_0-x)
\label{eq:delffit}
\end{align}
inspired by earlier work by Kulagin and Petti on off-shell PDF deformations in heavier nuclei \cite{Kulagin:2004ie}. The $x_0$ crossing and $\cal N$ normalization parameters are simultaneously fitted with the PDF and HT parameters, and $x_1$ is determined by requiring  that the off-shell correction does not modify the number of valence quarks in the nucleon,
\begin{align}
    \int_0^1 dx\, \delta f(x)\, \left[ q(x)-\bar q(x) \right] &= 0 
\label{eq:norm}
\end{align}
with $q=u,d$, see \cite{Accardi:2016qay} for details. More flexible parametrizations have been studied \cite{Li-Paris-2021} and will be reported elsewhere.
Finally, for ease of discussion, we collect the off-shell parameters into the vector
\begin{align}
    \vc a_\text{off} = ({\cal N},x_0,x_1) \ .
\end{align}

\subsection{Treatment of uncertainties}

The full set of fit parameters, including the PDF parameters discussed in Section~\ref{sec:PDF_par}, the higher-twist parameters and the off-shell parameters reads
\begin{align}
    \vc a = (\vc a_\textsc{pdf},\vc a_\textsc{ht},\vc a_\text{off} ) 
\end{align}
for a total number $n_\text{par}$ of parameters. The observables $\sigma$ we are interested in (for example the PDFs themselves, or the DIS structure functions, or the lepton pair production cross section) depend on the fitting parameters via the PDFs $f$, the HT function $C$, and the off-shell function $\delta f$. Schematically, 
\begin{align}
    \sigma [\vc a] = \sigma \big(f[\vc a_\textsc{pdf}],C[\vc a_\textsc{HT}],\delta f[\vc a_\text{off}] \big) \, .
\end{align}
The uncertainty on these observables can be estimated in the Hessian formalism~\cite{Pumplin:2001ct, Martin:2002aw}. With a sufficiently precise data set $\vc m = \{m_1, \ldots, m_{n_{dat}}\}$ and a suitably defined $\chi^2=\chi^2(\vc{a},\vc{m})$ chi-squared function, this method can approximate the parameter likelihood $\mathcal{L}(\vc{a}|\vc{m}) = \exp\big( -\frac{1}{2}\chi^2(\vc{a},\vc{m}) \big)$ as a multi-variate Gaussian distribution in parameter space centered around the best-fit value, $\vc{a_0}$, of the parameters~\cite{Hunt-Smith:2022ugn}. Namely, 
\begin{align}
    \mathcal{L}(\vc{a}|\vc{m})
    \propto \exp\Big( -\frac{1}{2}\Delta \vc{a}^T\, H\, \Delta \vc{a} \Big), 
\label{eq:Hess_approx}
\end{align}
where $m$ represent the data set being fitted, $\Delta\vc{a} = \vc{a}-\vc{a_0}$, and the Hessian matrix elements are given by 
\begin{equation}
    H_{ij} 
    = \frac12
    \left.
    \frac{\partial^2\chi^2(\vc{a})}{\partial a^i \partial a^j}
    \right|_{\vc a=\vc{a_0} }\,,
    \quad
    i,j=1,\dots n_{\rm par}\,.
\label{eq:hessmatrix}
\end{equation} 
The Hessian matrix can then be diagonalized, and reparametrized in terms of the eigendirections of the Hessian matrix via
\begin{equation}
    \vc{a}(\vc{t})=\vc{a_0}+\sum_{k=1}^{n_{\rm par}} t_k\frac{\vc{e_k}}{\sqrt{w_k}}\,,
\label{eq:parvar}
\end{equation}
where $\vc{e_k}$ and $w_k$ are the orthonormal eigenvectors and eigenvalues of the Hessian matrix, respectively, and $\vc t = \{t_1, \ldots, t_{n_{par}}\}$ is a vector of scaling factors. In terms of these variables, the approximated likelihood \eqref{eq:Hess_approx} is a symmetric Gaussian with $\vc t^2 =1$ identifying the 68\% confidence level on the fitted parameters. 

The standard CJ PDF error sets are then obtained by uniformly scaling each eigenvector by a ``tolerance factor'' $t_k = T$ to nominally produce an increase of $T^2$ above the minimum in the $\chi^2$ function. In both the \texttt{CJ15} and the \texttt{CJ15-a} analyses $T=1.645$ was chosen, corresponding to a 90\% Gaussian confidence level. In other analyses different choices are made to also account for tensions between the chosen data sets: for example, $T=10$ in the \texttt{CT10} global fit \cite{Lai:2010vv}.

However, in global QCD fits including \texttt{CJ15}, a few Hessian eigenvectors are typically not constrained enough by the available data and the likelihood can deviate from a Gaussian shape even within $t_k$ variations of order $O(1)$. This can happen, in particular: when data are scarce for a particular flavor combination, such as for $\bar d/\bar u$ at $x\gtrsim 0.3$; or closer to a kinematic threshold, such as for the $d/u$ ratio as $x \to 1$, where one expects the ratio to decrease towards 0, but not necessarily reaching that value \cite{Holt:2010vj}. 

A better approximation to the likelihood function \cite{Accardi:2021ysh,Hunt-Smith:2022ugn} can be obtained by scanning the $\chi^2$ function along each eigenvector starting from the best-fit parameters $\vc{a_0}$, until parameters $\vc{a_i}$ are found in the plus- and minus-directions such that the $\chi^2$ function increases above its best-fit value by an amount $T^2$:
\begin{align}
    \Delta\chi^2(\vc{a_{2i+1}}) &= \Delta\chi^2(\vc{a_{2i}}) = T^2 \nonumber \\
    &\hspace*{2cm} \forall\ i = 1, \ldots, n_{par} \ ,
\end{align}
where $\Delta \chi^2(\vc{a}) = \chi^2(\vc{a}) - \chi^2(\vc{a_0})$.
These parameter vectors correspond to a set of $t_k^\pm$ values that are close to $T$ wherever the Gaussian approximation holds, but can substantially deviate from this value along a few eigendirections. In other words, we adopt a local and asymmetric tolerance criterion instead of assuming $t_k=T$ globally. In practice, this scheme deforms the Hessian approximation of the likelihood in order to account in an approximate way for departures from a purely Gaussian behavior. It is also suitable for large $T$ values, for which the Hessian approximation cannot be \textit{a priori} assumed to hold. As with other global QCD analyses using local tolerance criteria \cite{Hou:2019efy,Bailey:2020ooq},
the price to be paid is that, while the chosen $T$ value legitimately defines a confidence region in parameter space, this cannot be readily and unambiguously associated with a confidence level figure as one can do with the pure Hessian approximation. 

In practical terms, we define parameter sets corresponding to variations along each eigendirection in the plus and minus directions, respectively, as 
\begin{align}
    \vc{a_{2k}} & = \vc{a_0} + t_k^+ \frac{\vc{e_k}}{\sqrt{w_k}} \\ 
    \vc{a_{2k+1}} & = \vc{a_0} - t_k^- \frac{\vc{e_{k}}}{\sqrt{w_k}} \ ,   
\end{align}
such that $\Delta \chi^2[\vc{a_i}] = T$ for all $i=1,\ldots,2n_{par}$. Then, the upper and lower $\delta \sigma_+$ and $\delta \sigma_-$ uncertainties on an observable $\sigma$ can be calculated using the expressions
\begin{subequations}
\begin{align}
\delta \sigma_+^2
& = \sum_{i=1}^{n_\text{par}}
      \Big[ \max\Big( \sigma[\vc{a_{2i-1}}]-\sigma[\vc{a_0}],
		      \sigma[\vc{a_2i}]-\sigma(\vc{a_0}),
			0
		\Big)
      \Big]^2
\label{eq:asym1}			\\
\delta \sigma_-^2
& = \sum_{i=1}^{n_\text{par}}
      \Big[] \max\Big( \sigma[\vc{a_0}]-\sigma[\vc{a_{2i-1}}],
		      \sigma[\vc{a_0}]-\sigma[\vc{a_{2i}}],
			0
		\Big)
      \Big]^2 \!\! .
\label{eq:asym2}
\end{align}
\end{subequations}
Alternatively, a symmetrized uncertainty can be obtained via
\begin{equation}
\delta \sigma^2
= \frac{1}{4}
  \sum_{i=1}^{n_\text{par}}
    \Big( \sigma[\vc{a_{2i-1}}] - \sigma[\vc{a_{2i}}] \Big)^2 \, .
\label{eq:sym_er}
\end{equation}
Note that the choice of tolerance $T$ value ($T=1.645$ in this paper) is already incorporated in Eqs.~\eqref{eq:asym1}-\eqref{eq:asym2} and \eqref{eq:sym_er}. The effect of alternative $T'$ tolerance choices can be approximately obtained by rescaling these uncertainties by a $T'/T$ factor as long as the two tolerance values are not too different from each other. However, care must be exercised for observables sensitive to the non-Gaussian regions of the parameter space.

\section{Dataset}
\label{sec:dataset}

\begin{table}[ht!]
\begin{tabular}{l|l|c|S|S}
    \hline
    Obs. & Experiment & Ref. & {\# Points}  & {$\chi^2$}\\
    \hline
     DIS & JLab (p) & \cite{JeffersonLabE00-115:2009jll}   & 136 & 161.0\\
         & JLab (d) & \cite{JeffersonLabE00-115:2009jll}   & 136 & 119.1\\
         & JLab (n/d) & \cite{CLAS:2014jvt} & 191 & 213.2\\
         & HERMES (p) & \cite{Hermes:2011} & 37  & 29.1\\ 
         & HERMES (d) & \cite{Hermes:2011} & 37  & 29.5\\ 
         & SLAC (p) & \cite{Whitlow:1991uw}   & 564 & 469.8\\
         & SLAC (d) & \cite{Whitlow:1991uw}  & 582 & 412.1\\
         & BCDMS (p) & \cite{BCDMS:1989qop}  & 351 & 472.2\\
         & BCDNS (d) & \cite{BCDMS:1990}  & 254 & 321.8\\
         & NMC (p) & \cite{NewMuon:1996fwh}    & 275 & 416.5\\
         & NMC (d/p) & \cite{NewMuon:1996uwk}  & 189 & 199.6\\
         & HERA (NC $e^{-}p)$ & \cite{H1:2015ubc} & 159 & 249.7\\
         & HERA (NC $e^{+}p$ 1) & \cite{H1:2015ubc} & 402 & 598.9\\
         & HERA (NC $e^{+}p$ 2) & \cite{H1:2015ubc} & 75 & 98.8\\
         & HERA (NC $e^{+}p$ 3) & \cite{H1:2015ubc} & 259 & 250.0\\
         & HERA (NC $e^{+}p$ 4) & \cite{H1:2015ubc} & 209 & 229.1\\
         & HERA (CC $e^{-}p)$ & \cite{H1:2015ubc} & 42  & 45.6\\
         & HERA (CC $e^{+}p)$ & \cite{H1:2015ubc} & 39  & 52.5\\
     LPP & E866 ($pp$) & \cite{NuSea:1998kqi} & 121 & 144.1\\
               & E866 ($pd$) & \cite{NuSea:1998kqi} & 129 & 157.4\\
               & SeaQuest ($d/p$) & \cite{SeaQuest:2021zxb} & 6 & 7.5\\
     W   & CDF ($e$) & \cite{CDF:2005cgc}   & 11 & 12.6\\
         & D0 ($e$) & \cite{D0:2014kma}     & 13 & 28.8\\
         & D0 ($\mu$) & \cite{D0:2013xqc}   & 10 & 17.5\\
         & CDF ($W$) & \cite{CDF:2009cjw}   & 13 & 18.0\\ 
         & D0 ($W$) & \cite{D0:2013lql}     & 14 & 14.5\\
         & STAR ($e^+/e^-$) & \cite{STAR:2020vuq} & 9  & 25.3\\
         & \hfill (less $\eta_\text{max}$ point) & & {(}8{\!\!\!\!\!)}  & {(}15.4{\!\!\!)}\\
     Z   & CDF & \cite{CDF:2010vek}       & 28 & 29.2\\
         & D0 & \cite{D0:2007djv}         & 28 & 16.1\\
     jet & CDF & \cite{CDF:2008hmn}       & 72 & 14.0\\
         & D0 & \cite{D0:2000dzr,D0:2008nou}         & 110& 14.0\\ 
     $\gamma$+jet & D0 1 & \cite{D0:2008chx} & 16 & 8.7\\
                  & D0 2 & \cite{D0:2008chx} & 16 & 19.3\\
                  & D0 3 & \cite{D0:2008chx} & 12 & 25.0\\
                  & D0 4 & \cite{D0:2008chx} & 12 & 12.2\\
    \hline
    & total         & & 4557 & 4936.6\\
    & total + norm  & & 4573 & 4948.6\\
    \hline
\end{tabular}
\caption{\label{tab1:data} Data sets and corresponding number of data points and $\chi^2$ values from the \texttt{CJ22} analysis.}
\end{table}

\begin{figure*}[hbt!]
    \centering
    \includegraphics[width=0.49\linewidth]{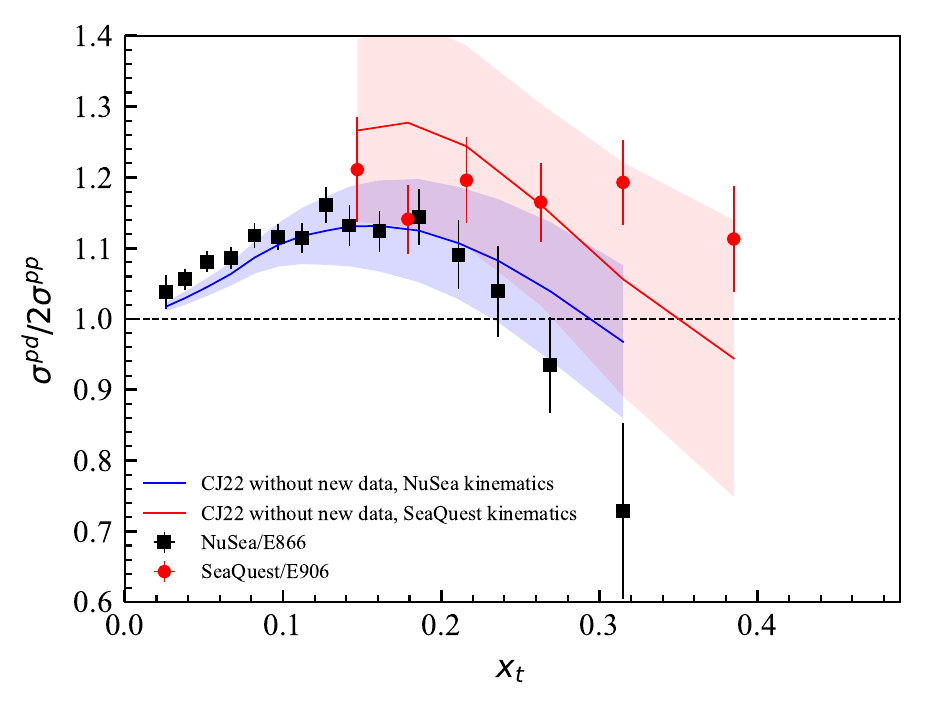}
    \includegraphics[width=0.49\linewidth]{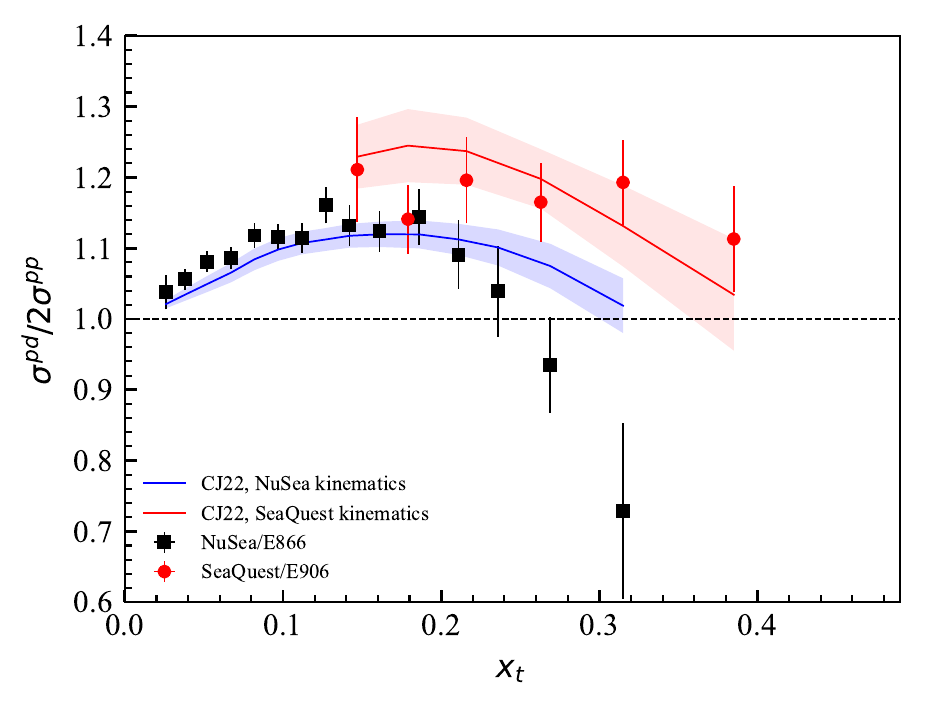}
    \caption{\label{fig:DY} Comparison of the measured cross section ratio for lepton pair production in $pd$ and $pp$ collisions from the E866~\cite{NuSea:1998kqi,NuSea:2001idv} and SeaQuest~\cite{SeaQuest:2021zxb} experiments with NLO calculations. The solid blue (red) curve with $T=1.645$ uncertainty band represents the ratio calculated at the E866 (SeaQuest) kinematics before (left) and after (right) including the new weak boson production data from SeaQuest and STAR in the fit.}
\end{figure*}

The \texttt{CJ15} analysis included DIS data from fixed target electron-hadron scattering experiments at Jefferson Lab~\cite{JeffersonLabE00-115:2009jll, CLAS:2014jvt}, HERMES~\cite{Hermes:2011}, SLAC~\cite{Whitlow:1991uw}, BCDMS~\cite{BCDMS:1989qop,BCDMS:1990}, and NMC~\cite{NewMuon:1996fwh,NewMuon:1996uwk}, and from the HERA $ep$ collider~\cite{H1:2015ubc}; $W$~\cite{CDF:2005cgc,D0:2014kma,D0:2013xqc,CDF:2009cjw,D0:2013lql} and $Z$~\cite{CDF:2010vek,D0:2007djv} asymmetries as well as jet~\cite{CDF:2008hmn,D0:2000dzr,D0:2008nou} and $\gamma$+jet~\cite{D0:2008chx} data from CDF and D0 experiments at Tevatron; lepton pair production (LPP) from the E866 experiment at Fermilab \cite{NuSea:1998kqi}.

Previously, the antiquark PDFs in the mid-$x$ region were mainly constrained by the lepton pair production data from the E866 experiment. In the \texttt{CJ22} fits we have now included recent data that are sensitive to the antiquarks from the new lepton pair production measurements by the E906/SeaQuest experiment~\cite{SeaQuest:2021zxb} and the rapidity distribution of the $W^+/W^-$ ratio in $pp$ collisions by the STAR experiment~\cite{STAR:2020vuq}.
The data sets used in the \texttt{CJ22} fit are listed in Table~\ref{tab1:data}.

The SeaQuest data covers the kinematic range $0.1 < x < 0.45$ extending the large $x$ reach from the E866 experiment at different $Q^2$. 
The cross section ratio of the lepton pair production in $pp$ and $pd$ interactions attracts particular interest as it can be directly related to $\bar{d}/\bar{u}$. In the forward region, the ratio can be written as
\begin{equation}
    \frac{\sigma_{pd}}{\sigma_{pp}} 
    \approx \frac{4+\frac{d(x_b)}{u(x_b)}}
    {4+\frac{d(x_b)}{u(x_b)} \frac{\bar d(x_t)}{\bar u(x_t))}} 
    \left( 1 + \frac{\bar{d}(x_t)}{\bar{u}(x_t)} \right) \ .
\label{eq:DYratio}
\end{equation}
When $x_b \to 1$, the $d/u$ ratio tends to 0 and can be neglected, so that the cross section ratio becomes sensitive only to the $\bar d/\bar u$ ratio. However, neither for the E866 experiment ($x_b = 0.3 - 0.5$) nor for the SeaQuest experiment ($x_b = 0.5 - 0.7$) is this condition satisfied, and the data are sensitive both to the $\bar d/\bar u$ quark ratio and, subdominantly, to the $d/u$ ratio.

In $pp$ collisions, $W$ bosons are produced from quark-antiquark fusion and therefore provide clean access to quark and antiquark distributions inside the proton at a large momentum scale $Q^2 = M_{W}^2$. The STAR experiment at RHIC has recently reported the unpolarized $W$ and $Z$ boson cross sections at $\sqrt{s}$ = 510 GeV via $W^{\pm}\rightarrow e^{\pm}+\nu$ and $Z\rightarrow e^{+}e^{-}$ decays, respectively in the pseudorapidity range $-1.0 < \eta < 1.5$. An observable that is particularly sensitive to  the $\bar{d}(x)/\bar{u}(x)$ ratio is the $W^{+}/W^{-}$ ratio of $W$ boson cross sections. At leading order this can be written as
\begin{equation}
    \frac{\sigma_{W^{+}}}{\sigma_{W^{-}}} \ \approx \ \frac{u(x_1)\bar{d}(x_2)+\bar{d}(x_1)u(x_2)}{d(x_1)\bar{u}(x_2)+\bar{u}(x_1)d(x_2)}
   \ \underset{y_{_W} \approx 0}{\approx} \ \frac{\bar d / \bar u}{d / u }
   \label{eq:Wratio}
\end{equation}
where $x_{1,2} = (M_{_W}/\sqrt{s}) \exp(\pm y_{_W})$ is the fractional momentum carried by the scattering partons, with $y_{_W}$ the rapidity of the produced boson. At midrapidity, where $x_1=x_2\approx 0.16$, the cross section ratio directly accesses both the antiquark and the quark ratios. 
At larger rapidity, the accessible $x_{1,2}$ range is somewhat limited by the boson decay kinematics as well as by the statistical precision of the data, and the measured lepton asymmetry effectively probes light quarks and antiquarks with fractional momenta $x$ in the $0.05 \lesssim x \lesssim 0.25$ range.

Of the STAR measurements, the $W$ boson charge ratio is the most sensitive to the quark and antiquark ratios, and has been included in the \texttt{CJ22} analysis. We have not included either of the charged separated $W$ or the $Z$ measurements because these do not provide significant additional constraints on the PDF determination, but we will discuss in the next section how well the new fit describes those data. 
As already mentioned, for the STAR $W$ and $Z$ cross section calculations we use fast NLO interpolation grids that were created using APPLgrid \cite{Carli:2010rw} interfaced with the MCFM event generator \cite{Campbell:1999ah,Campbell:2012uf}.
The events were generated using the experimental cuts for electron transverse momentum ($p_e > 15$ GeV/$c$) and energy ($25 < E_e < 50$ GeV). The STAR $W^{\pm}\rightarrow e^{\pm} + \nu$ measurements also require cuts to suppress jet background. To reproduce a similar condition, we excluded events with produced jets, obtaining approximately a 20\% reduction in the calculated cross section. For $Z \rightarrow e^+e^-$ events, the generated events are collected within the experimental invariant mass range of 70 GeV $< M_{e^+e^-} < $ 110 GeV for electron pairs. During the fit, the grids so obtained are then convoluted with the PDFs to calculate the needed NLO cross sections.

\section{Results}
\label{sec:results}

\begin{figure*}[thb]
    \begin{subfigure}
        \centering
        \includegraphics[width=0.35\linewidth]{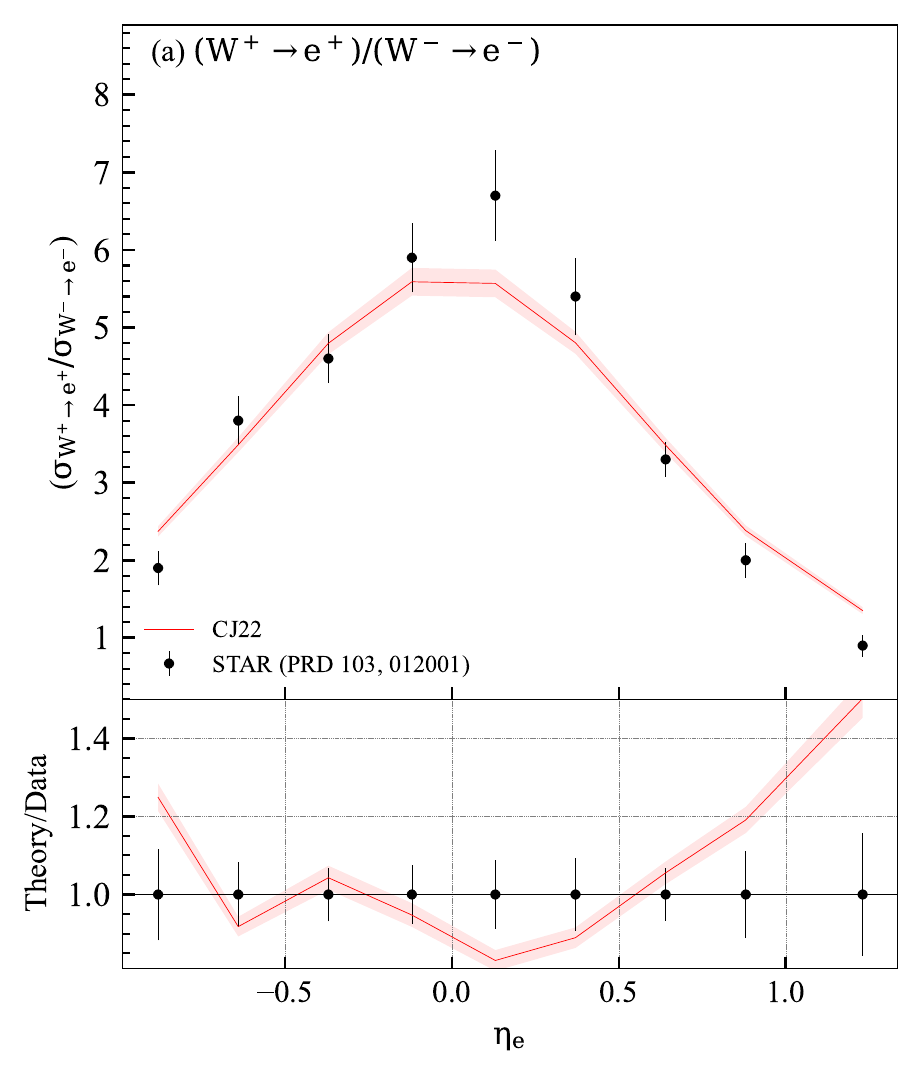}
    \end{subfigure}
    \begin{subfigure}
        \centering
        \includegraphics[width=0.35\linewidth]{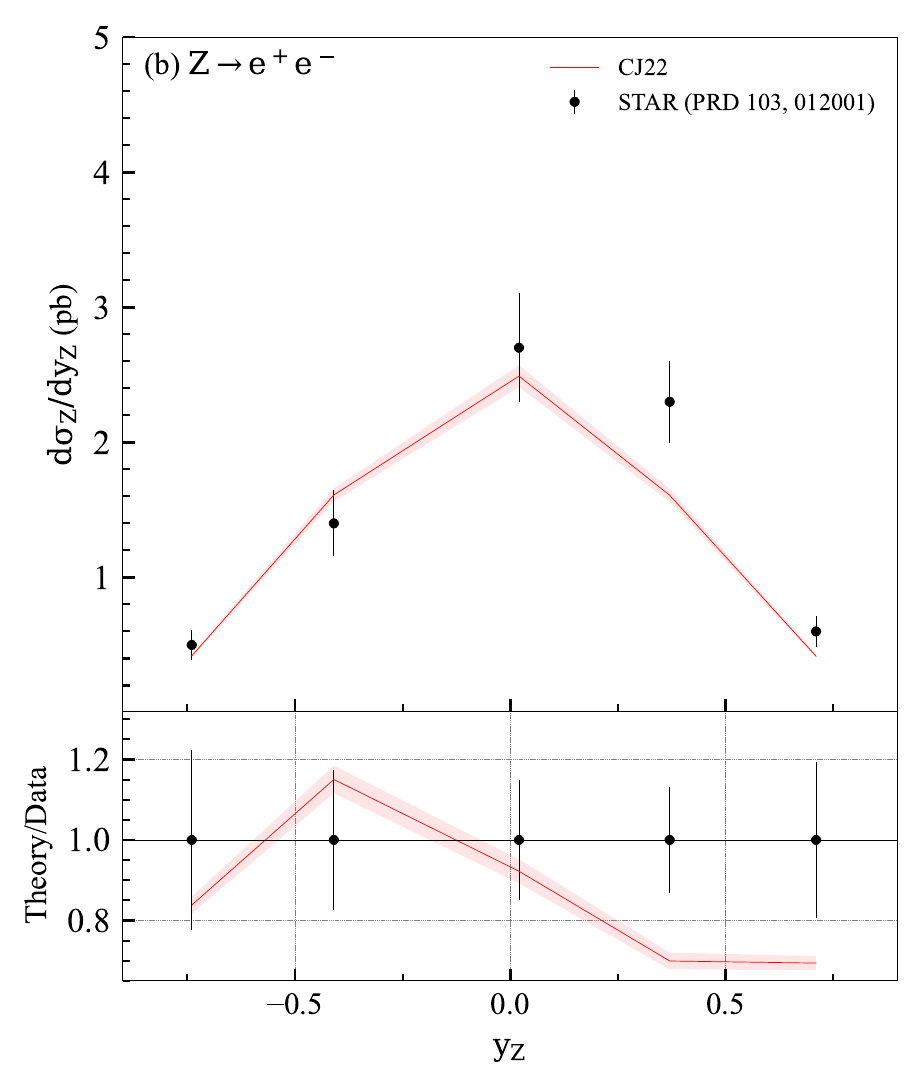}
    \end{subfigure}
    \begin{subfigure}
        \centering
        \includegraphics[width=0.35\linewidth]{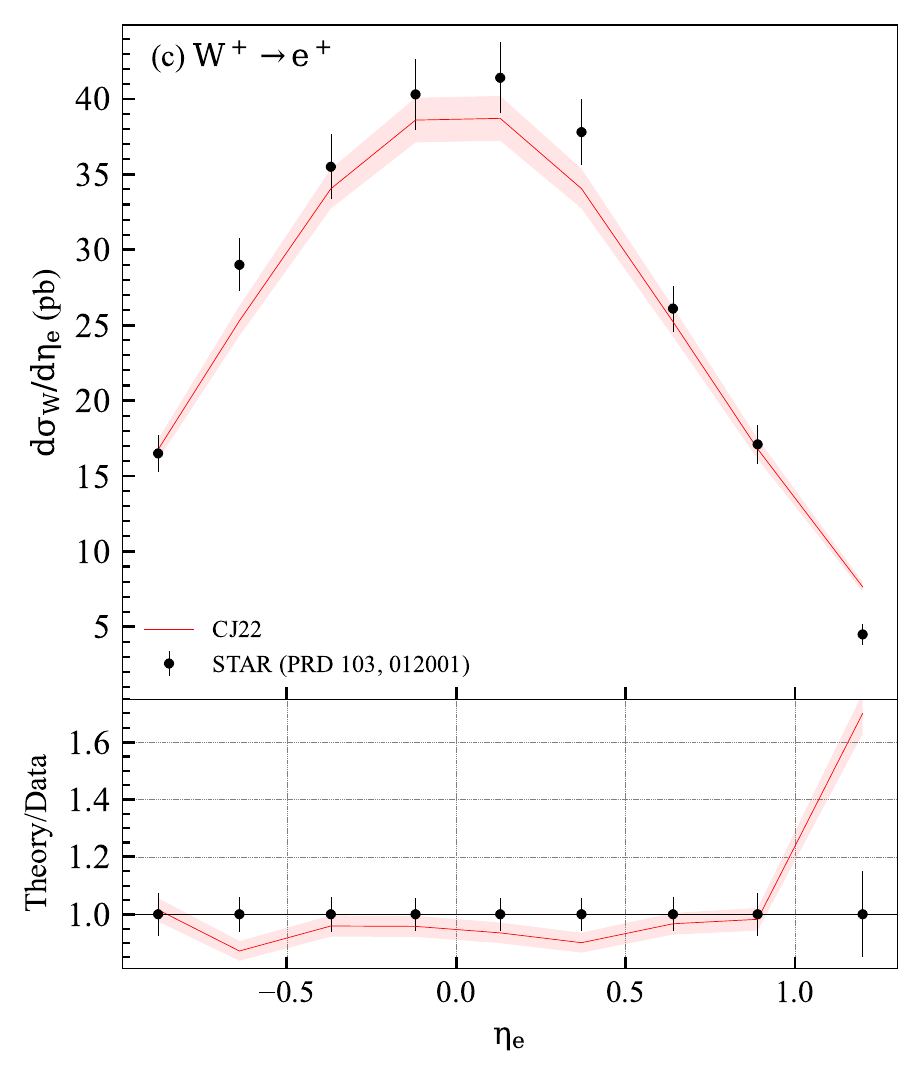}
    \end{subfigure}
    \begin{subfigure}
        \centering
        \includegraphics[width=0.35\linewidth]{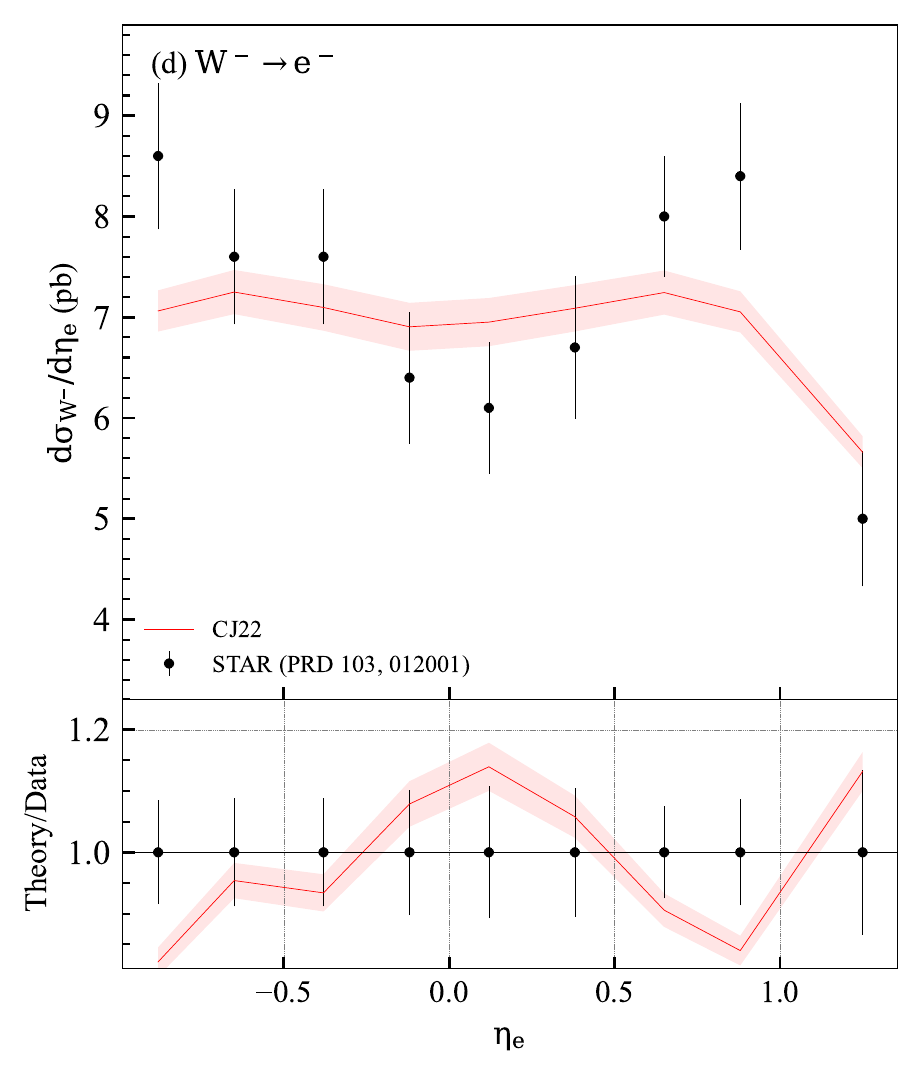}
    \end{subfigure}
    \caption{\label{fig:obs_STAR}The measured (a) $\sigma_{W^{+}}/\sigma_{W^{-}}$, (b) $d\sigma_{Z}/dy_{Z}$, (c) $d\sigma_{W^{+}}/d\eta_e^{+}$ and (d) $d\sigma_{W^{-}}/d\eta_e^{-}$ are compared with the \texttt{CJ22} calculations. The statistical and the total systematic uncertainties are added in quadrature and shown as the solid error bars for the data points. The solid red lines show the central values from our fit. The red bands correspond to the $T=1.645$ PDF uncertainty. The differences with respect to \text{CJ15} calculations are minor and the corresponding curves are omitted for visual clarity.
    }
\end{figure*}

\begin{figure*}[thb]
    \centering
    \includegraphics[width=1.0\linewidth]{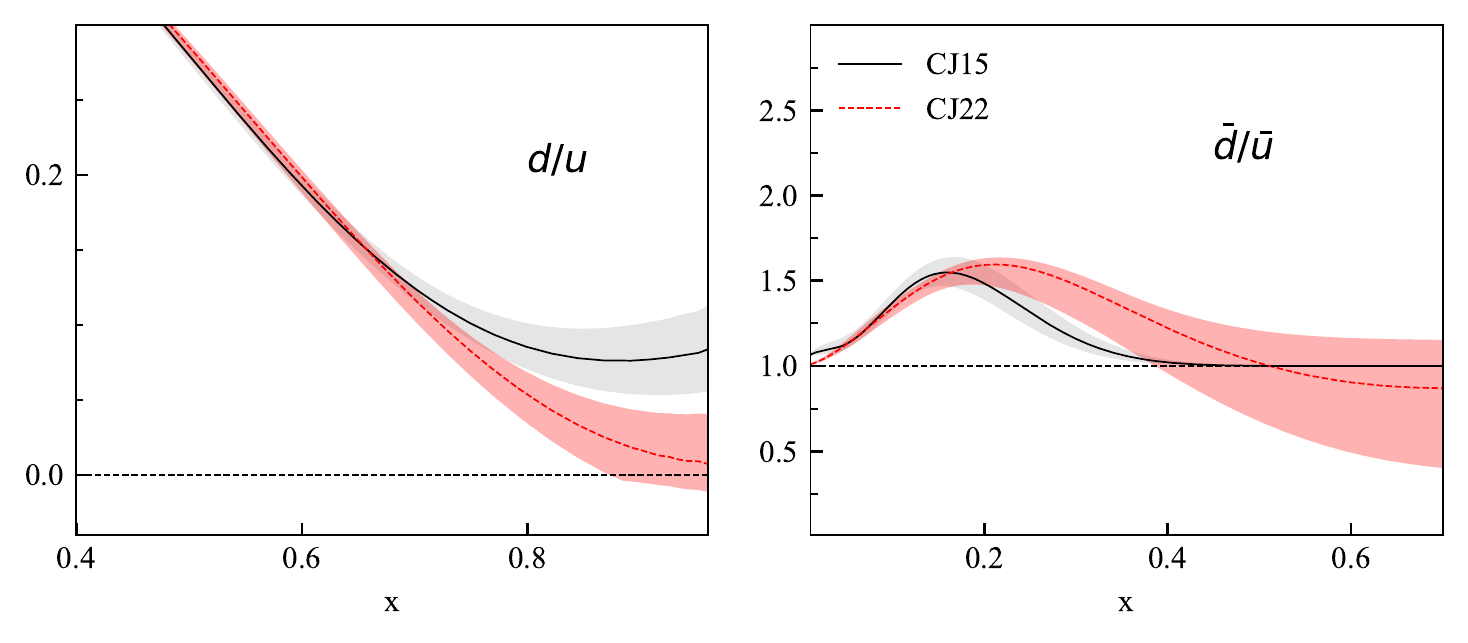}
    \caption{\label{fig:dbub} Comparison of the $d/u$ (left) and $\bar{d}/\bar{u}$ (right) at $Q^2$ = 10 GeV$^2$ between \texttt{CJ15} (black solid curve) and \texttt{CJ22} (red dashed curve) with 90\% CL uncertainty bands.}
\end{figure*}

In this Section, the results of the new analysis are presented and compared with the previously published \texttt{CJ15} results \cite{Accardi:2016qay}. 

Figure~\ref{fig:DY} compares the lepton pair production cross section ratios from E866 and SeaQuest with our calculations before and after including SeaQuest data into the \texttt{CJ22} fit. The comparison is done separately for the kinematics of each experiment, with E866 using a 800 GeV proton beam and SeaQuest a 120 GeV beam, and each experiment accessing a different range of lepton pair mass $M$. For the SeaQuest data, we include the spectrometer acceptance matrix provided in Ref.~\cite{SeaQuest:2021zxb}, which has a relatively small effect on an already relatively flat observable.

Looking at Eq.~\eqref{eq:DYratio}, if $d(x_b)/u(x_b) \ll 1$ was really negligible, as is often assumed for simplicity, one would expect that the ratio of the proton to deuteron cross section data should be approximately independent of $M^2$. Instead, the ratio measured by the experiments is different, which is a direct reflection of the role of the $d(x_b)/u(x_b)$ ratio in Eq.~\eqref{eq:DYratio}. Indeed, at the higher $x_b$ probed by SeaQuest that ratio is smaller than at the lower $x_b$ values of the corresponding E866 measurement. Therefore, one should expect the cross section ratio to be higher for SeaQuest than for E866, which is confirmed in Figure~\ref{fig:DY}. Furthermore, Eq.~\eqref{eq:DYratio} shows that an increase in the antiquark $\bar d(x_t)/\bar u(x_t)$ ratio can be compensated in the PDF fit by a decrease in the $d(x_b)/u(x_b)$ quark ratio and vice versa. The resulting anticorrelation will be important to understand the behavior of the fitted PDF ratios that will be discussed later.

In the left panel of Figure~\ref{fig:DY}, the calculations from the fit that only includes the E866 data show a steeper downturn in the cross section ratio than allowed by the SeaQuest data. With the new data added to the fit, however, the \texttt{CJ22} PDFs bring the ratio plotted in the right panel distinctly above 1 in the large-$x$ region where E866 has limited kinematic coverage and statistical precision, with substantially reduced PDF uncertainty. While the new cross section calculation lies higher than the last two E866 data points, only a minor increase in $\chi^2$/datum from 1.63 to 1.93 is observed for the E866 data because of the relatively large uncertainty of the last few data points. Conversely the $\chi^2$/datum value sharply reduces from 3.19 to 1.25 for the SeaQuest data after including the new data, reflecting both the enhanced kinematic range and the precision of the new data.

The CJ22 fit also included the $W^+ \!\to e^+ \!/ W^- \!\to e^-$ cross section ratio measured by the STAR collaboration, which, as discussed in the previous Section, provide complementary information on the $\bar{d}/\bar{u}$ ratio around a smaller $x \approx 0.16$ value, overlapping with the E866 data but at a higher scale.
The quality of the fit to the charge ratio data is shown in the upper left panel of Fig.~\ref{fig:obs_STAR}. The remaining panels show a comparison of NLO calculations using the new \texttt{CJ22} PDFs to the unfitted STAR data on the $Z$, $W^+ \!\to e^+$ and $W^- \!\to e^-$ rapidity distributions. Differences with \texttt{CJ15} calculations are minor, and the corresponding curves not shown in the plots.

Overall, \texttt{CJ22} describes reasonably well the $W$ and $Z$ measurements. However, there is a suggestion of more structure in the $W^- \!\rightarrow e^-$ channel than shown by the theory. One can also note that the highest rapidity point of the $W^+$ cross section is much lower than the corresponding theoretical calculation, which has PDFs strongly constrained by the rest of the world data set and cannot accommodate such a small measurement. Similar features were also observed in other PDF analyses of the STAR data~\cite{STAR:2020vuq,Cocuzza:2021cbi,Guzzi:2021fre}.
In fact, reducing the calculated $W^+$ cross section in this region would require substantially increasing the $d(x_1)/u(x_1)$ ratio at large values of $x_1$ and/or decreasing the value of the $\bar d(x_2)/\bar u(x_2)$ ratio at small values of $x_2$. Both possibilities would cause the fits to the lepton pair production data to be worse. Hence, it has not proven possible to get a good description of this one data point. 

Figure~\ref{fig:dbub} shows the impact of the new data on $d/u$ and $\bar{d}/\bar{u}$ ratios at a scale of $Q^2$ = 10 GeV$^2$. The results are compared with the \texttt{CJ15} light quark and antiquark ratios. In the \texttt{CJ15} analysis, the E866 data provided the strongest constraints for the light-antiquarks and the larger-$x$ region ($x > 0.3$) was essentially left unconstrained by data. As a result, the \texttt{CJ15} analysis was performed with a more rigid parametrization of the light-antiquarks, and $\bar{d}/\bar{u}$ ratio was forced to approach 1 as $x\rightarrow 1$. The new data from SeaQuest adds significant constraints on the $\bar{d}/\bar{u}$ with a larger reach in $x$ and allowed us to relax the parametrization used in \texttt{CJ22}, which does not prescribe the large $x$ behavior of the $\bar{d}/\bar{u}$ ratio. The \texttt{CJ22} fit obtains a $\bar{d}/\bar{u}$ ratio that keeps increasing until $x \approx 0.25$ in the region where E866 data would have required a sharp drop. At $x \gtrsim 0.25$, the ratio naturally starts decreasing, but remains above 1 within uncertainties, as is also the case for the $\bar d/\bar u$ ratio obtained by the  CT \cite{Guzzi:2021fre} and JAM \cite{Cocuzza:2021cbi} collaborations after inclusion of the SeaQuest data in their global fit. At $x\lesssim 0.2$ the antiquark ratio is driven by the STAR data slightly below the \texttt{CJ15} result, but remains compatible with the latter. 
Turning to the light quark ratio displayed in the left panel of Figure~\ref{fig:dbub}, one can see that the \texttt{CJ15} $d/u$ ratio remains decidedly above 0 at large $x$, with a central value extrapolated to $x = 1$ of $0.09 \pm 0.03$. On the contrary, in the \texttt{CJ22} analysis, the ratio approaches 0 within uncertainties as $x \rightarrow 1$. This is due to the anticorrelation between the $\bar{d}/\bar{u}$ and $d/u$ induced by the lepton pair production data, and evidenced in Eq.~\eqref{eq:DYratio}, that was discussed earlier:
the increase of $\bar{d}/\bar{u}$ in the medium $x$ region, which is allowed by the more flexible \texttt{CJ15-a} and  \texttt{CJ22} parametrizations, and further driven by the increased kinematic reach of the SeaQuest data in the latter, has caused a decrease in $d/u$ in the large $x$ region. The \texttt{CJ15} non-zero $d/u$ limit was thus the result of a parametrization bias that also underestimated the nominal uncertainty band. With this bias removed, the new result is compatible with the recent $d/u$ fits performed by Al\"ekhin, Kulagin and Petti \cite{Alekhin:2017fpf, Alekhin:2022tip}, that have a similar large-$x$ theoretical setup and data coverage (except for the SeaQuest and STAR data).
\section{Summary}
\label{sec:conclusions}

We have presented the results of our recent \texttt{CJ122} global QCD analysis of parton distributions which included new electroweak data from SeaQuest and STAR. The SeaQuest  data, in particular, extends the $x$ coverage to larger $x < 0.45$ compared to the previous measurement by E866, leading to significant constraints on the $\bar{d}/\bar{u}$ ratio, allowing the use of the more flexible light-antiquark parametrization discussed in the text. In the \texttt{CJ22} fit the $\bar d/\bar u$ ratio remains near 1 as $x \rightarrow 1$ without having to build that into its parametrization, as was previously done in \texttt{CJ15}. The data are also sensitive to the $d/u$ light quark ratio, and the interplay between this and $\bar{d}/\bar{u}$ leads to a $d/u$ ratio that lies below that found in \texttt{CJ15} as $x \rightarrow 1$, and is compatible with 0 in that limit. 
\section*{Acknowledgments}  

We would like to thank J. Bane, S. Fazio, M. Posik, and A. Tadepalli for informative discussions on their experimental measurements, as well as C. Cocuzza and W. Melnitchouk for useful comments and criticism.

This work was supported in part by the  U.S. Department of Energy (DOE) contract DE-AC05-06OR23177, under which Jefferson Science Associates LLC manages and operates Jefferson Lab. 
AA also acknowledges support from DOE contract DE-SC0008791.
X. Jing was partially supported by DOE Grant No. DE-SC0010129.
S. Park acknowledges support from DOE contract DE-FG02-05ER41372 and the Center for Frontiers in Nuclear Science.

\bibliography{fullbib}   
\end{document}